# Coexistence of Magnetic Orders in Two-Dimensional Magnet CrI$_3$

Ben Niu,[1,2†] Tang Su,[3,2†] Brian A. Francisco,[2] Subhajit Ghosh,[4] Fariborz Kargar,[4] Xiong Huang,[5] Mark Lohmann,[2] Junxue Li,[2] Yadong Xu,[2] Takashi Taniguchi,[6] Kenji Watanabe,[6] Di Wu,[1] Alexander Balandin,[4] Jing Shi,[2*] Yong-Tao Cui[2*]

1. National Laboratory of Solid State Microstructures, Department of Materials Science and Engineering, Jiangsu Key Laboratory for Artificial Functional Materials, and Collaborative Innovation Center of Advanced Microstructures, Nanjing University, Nanjing 210093, China
2. Department of Physics and Astronomy, University of California, Riverside, CA 92521, USA
3. International Center for Quantum Materials, School of Physics, Peking University, Beijing 100871, P. R. China
4. Department of Electrical and Computer Engineering, University of California, Riverside, CA 92521, USA
5. Department of Materials Science and Engineering, University of California, Riverside, CA 92521, USA
6. National Institute for Materials Science, 1-1 Namiki, Tsukuba, 305-0044, Japan

† These authors contribute equally to this work.
* Email: jing.shi@ucr.edu, yongtao.cui@ucr.edu

**Abstract**

The magnetic properties in two-dimensional van der Waals materials depend sensitively on structure. CrI$_3$, as an example, has been recently demonstrated to exhibit distinct magnetic properties depending on the layer thickness and stacking order. Bulk CrI$_3$ is ferromagnetic (FM) with a Curie temperature of 61 K and a rhombohedral layer stacking, while few-layer CrI$_3$ has a layered antiferromagnetic (AFM) phase with a lower ordering temperature of 45 K and a monoclinic stacking. In this work, we use cryogenic magnetic force microscopy to investigate CrI$_3$ flakes in the intermediate thickness range (25 – 200 nm) and find that the two types of magnetic orders hence the stacking orders can coexist in the same flake, with a layer of ~13 nm at each surface being in the layered AFM phase similar to few-layer CrI$_3$ and the rest in the bulk FM phase. The switching of the bulk moment proceeds through a remnant state with nearly compensated magnetic moment along the c-axis, indicating formation of c-axis domains allowed by a weak interlayer coupling strength in the rhombohedral phase. Our results provide a comprehensive picture on the magnetism in CrI$_3$ and point to the possibility of engineering magnetic heterostructures within the same material.

The recent discoveries of intrinsic magnetism in several van der Waals materials[1,2] have generated widespread interest in the study of magnetism in two dimensions[3–5], as their vdW nature allows exfoliation of thin flakes down to the monolayer limit. These 2D magnetic systems exhibit exciting magnetic, electrical and optical properties, such as giant tunneling magnetoresistance,[6–9] magnetoelectric coupling,[10–12] and magnetic second harmonic generation,[13] promising for future spintronics applications. Meanwhile, several important questions regarding their magnetic properties in both bulk crystals and few layer flakes remain elusive. In bulk $CrI_3$ crystal, previous magnetic characterizations indicate a ferromagnetic order with a c-axis anisotropy and a small (<0.2 T) magnetic field is able to saturate the magnetization along the c-axis.[14,15] However, the magnetization vs field (M-H) loop shows little hysteresis and a very small remnant magnetization at zero field, which was explained by formation of magnetic domains in bulk crystals but yet to be confirmed.[14] On the other hand, exfoliated $CrI_3$ flakes in the few-layer limit show distinct magnetic properties from the bulk. Few layer $CrI_3$ is found to be a layered antiferromagnet with a Curie temperature of ~45 K compared to the bulk $T_c$ of ~61 K.[1,8,9,16] It also requires a much larger magnetic field (~1-2 T) to reverse individual layers to polarize the magnetization.[1,6–9] Recent experimental[13,17–19] and theoretical[20–23] studies have suggested that the difference in the magnetic order originates from different layer stacking orders: bulk $CrI_3$ undergoes a structural phase transition from its high-temperature monoclinic phase to a rhombohedral phase at ~200 K, but this transition is suppressed in exfoliated few-layer $CrI_3$ so that it remains in the monoclinic phase at low temperatures with a different magnetic configuration. It is thus interesting to examine over what thickness range the layer stacking order recovers to the bulk phase. To address these questions, we perform cryogenic magnetic force microscopy to study the magnetic properties in thin $CrI_3$ flakes of various thicknesses. We find that the two types of magnetic configurations coexist in thin $CrI_3$ flakes: surface layers of ~13 nm exhibit an AFM interlayer coupling similar to exfoliated few-layer $CrI_3$ in the monoclinic phase, while the inner layers behave similarly to $CrI_3$ bulk crystals with a FM interlayer coupling which is weaker in strength than that of the AFM coupling in the surface layers, resulting in a remnant state with nearly compensated magnetic moment along the c-axis.

In our experiment, thin $CrI_3$ flakes were prepared by mechanical exfoliation and covered by hexagonal boron nitride (hBN) flakes in an inert environment to minimize $CrI_3$ degradation. (See more details in Supporting Information (SI).) Magnetic force microscopy was performed in a cryogenic environment with a magnetic field applied perpendicular to $CrI_3$ flakes. In MFM measurement, a cantilever probe coated with a thin magnetic film was used to sense the force exerted by the sample's magnetic fields on the probe. Specifically, the change in the probe's mechanical resonant frequency is proportional to the spatial derivative of the magnetic field. Therefore, MFM is highly sensitive to the formation of lateral magnetic domains in which domain walls produce strong spatial variations of magnetic fields, but a laterally uniform magnetization also produces a finite but smaller MFM signal as the magnetic field decays away from the sample surface, as illustrated by a simulated MFM curve shown in Fig. 1a. In the latter case, the MFM signal is proportional to the areal density of the sample moment integrated over the thickness dimension, which can be used as a nanoscale magnetometer to measure the local magnetization strength.

First, we examine the magnetic configuration during the reversal process in a 200 nm $CrI_3$ flake. Figure 1b shows representative MFM images of this flake at different magnetic fields. At $B=5$ T where the magnetization is fully aligned, the sample shows a roughly uniform MFM signal inside the flake which only has strong variations near the $CrI_3$ edge due to the geometry of the stray field. The negative sign of the MFM signals in the interior of the flake indicates an attractive force because the moments of both sample and probe are aligned along the same direction by the applied magnetic field. The weak contrast pattern in the interior corresponds to regions of different thicknesses hence different magnetic moments. (See SI for a comparison with the sample topography.) The roughly uniform configuration persists as the field is reduced all the way to around 0.2 T, except within a narrow field range between 2.1 T and 1.9 T where magnetic domains appear. But the MFM signal inside the bulk only changes slightly through this range, indicating a partial reversal of the moments. Until about 0.2 T, the average MFM signal in the bulk starts to change quickly, and domains with weak contrasts appear and evolve continuously as the field is further reduced. Around -0.1 T, the average signal reaches around 0 while the signal contrast across domains remains weak. When the field is further reduced below -0.1 T through -5 T, MFM signal follows an evolution similar to the positive field side. (See SI for more MFM images.) Such weak MFM signal pattern suggests that although domains form during the reversal process, the thickness-integrated areal density of moments in individual domains are small, similar to the simulated case in Fig. 1a (ii). In the remnant state, the total moment is nearly compensated along the c-axis.

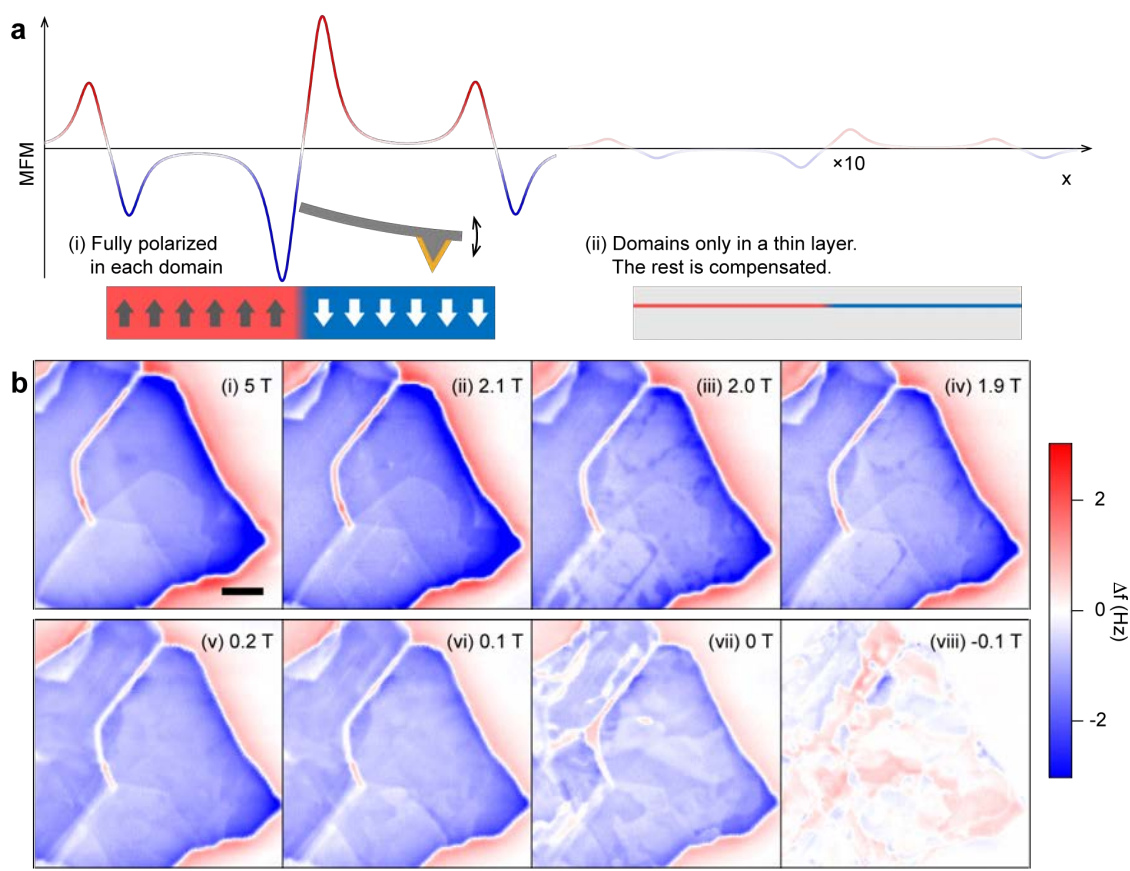

**Figure 1. MFM imaging of the magnetic switching in exfoliated $CrI_3$.** (a) Simulated MFM signal across magnetic domains in two scenarios: (i) Moments are fully polarized in each domain. (ii) Domains only exist in a thin layer with zero net moment in rest of the sample. The MFM signal in (ii) is scaled by a factor of 10. (b) MFM images of a 200 nm $CrI_3$ flake with magnetic field decreasing from 5 T to -0.1 T. Scale bar is 2 μm.

Next, we study more quantitatively the change in magnetic moment during the reversal process and demonstrate that two types of magnetic configurations coexist in CrI$_3$ flakes. The MFM images suggest that the magnetic configuration during most of the reversal process is spatially uniform. In this case the MFM signal will be roughly proportional to the areal density of magnetic moment integrated over the flake thickness. Based on this property, we measure the MFM signal inside the flake as a function of magnetic field, as shown in Fig. 2a. (Note that we plot -Δf to represent the magnitude of the magnetic moment where the negative sign accounts for the attractive nature of the magnetic force.) The MFM vs *B* curve captures the switching process as the magnetic moment is fully reversed, which exhibits two distinct stages as hinted already by the MFM images. When sweeping down from the fully polarized state at 5 T, the MFM signal remains roughly constant until around 2 T where it decreases by a step over a small field range. It then again remains roughly constant until about 0.2 T. Below 0.2 T, the signal quickly decreases to zero at -0.1 T. After the field is further swept to negative values, the behavior of the MFM signal is similar to the positive field side: the magnetization is reversed along the opposite direction in two stages. (Note that the moment of the MFM tip is also reversed at negative field, so the sign of the MFM signal remains the same.) The two stages of the switching process suggest that this 200 nm $CrI_3$ flake can be divided into two groups of layers with different magnetic properties. One group has its switching field at ~2 T (referred to below as "hard layers"), which is very similar to the switching properties in few-layer $CrI_3$. In contrast, the other group of "soft layers" switches at low field, similar to the soft switching behavior seen in $CrI_3$ bulk crystals. From the relative ratio of the signal changes for these two groups, we calculate the effective thickness of each group: the hard layers contribute ~12% of the full-scale signal which converts to ~24 nm in thickness, and the soft layers span the rest. This behavior of two-stage switching is quite general in flakes of different thicknesses. In all flakes that we have measured, there always exists one group of hard layers that switch at ~2 T. (See Fig. 2b and 2c and more examples in SI.) The relative percentage of this group increases as the flake thickness decreases, but the calculated effective thickness is roughly a constant around 25 nm which varies slightly in different samples (Fig. 2d). In the 25 nm flake, all the moments switch together at ~2 T, and no soft layers can be identified. This thickness dependence can be explained by that the hard layers correspond to the surface layers of the $CrI_3$ flake while the soft layers correspond to the inner layers. As the total thickness decreases, the thickness of the hard surface layers remains constant while the soft inner layer portion shrinks, until when the total thickness is below ~25 nm the soft inner layer group disappears and the entire flake behaves as one group that switch only at ~2 T. Since every flake always has two surfaces, the hard layers at each surface should have a thickness of ~13 nm.

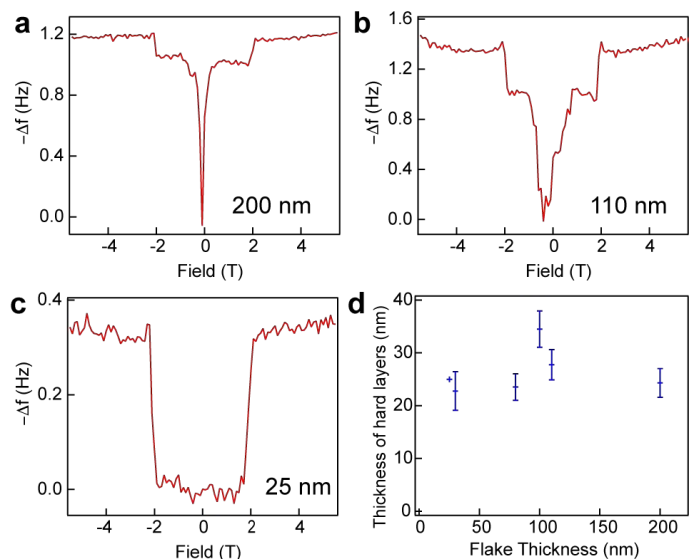

**Figure 2. Two groups of layers with different magnetic properties coexist in CrI$_3$ flakes.** (a-c) MFM signal as a function of magnetic field for three flakes with thicknesses of (a) 200 nm, (b) 110 nm, and (c) 25 nm. (d) Calculated thicknesses of the hard layers in CrI$_3$ flakes with thicknesses ranging from 25 to 200 nm.

The hard and soft layers also have different Curie temperatures. Figure 3a plots the MFM signal vs $B$ curves for the 200 nm flake at temperatures from 10 K up to 100 K. As temperature increases, the switching field of the hard layers decreases from 2 T and the associated change in the MFM signal decreases as well. Around 45-50 K, no switching can be identified for this group. Figure 3b plots the temperature dependence of the switching field and the MFM signal change for the hard layers, from which we estimate its Curie temperature to be ~45 K, similar to the Curie temperature in few layer $CrI_3$. In contrast, the soft switching at low field still persists up to 60 K. The size of the signal which reflects the magnetic moment of the soft inner layers decreases with increasing temperature, and reaches zero between 60-70 K, resembling the behavior in $CrI_3$ bulk crystal which has a Curie temperature of 61 K.

Based on our observations, we present the following picture to describe the magnetic configuration in these $CrI_3$ flakes, as illustrated in Fig. 3d. First, the surface layers have a layered AFM configuration which is similar to that in few-layer $CrI_3$, i.e., ferromagnetically coupled within each layer but antiferromagnetically coupled between adjacent layers.[1] These layers switch from the layered AFM state to the fully polarized FM state at ~ 2 T, as indicated by the abrupt change in the MFM signal. Second, during the switching of the inner layers at low fields, the gradual decrease of the MFM signal and the lack of strong domain contrasts suggests that the inner layers switch layer by layer where the weak lateral domain contrasts indicate the formation of lateral domains within the individual layer that is undergoing switching. At the remnant state, the CrI$_3$ flakes has a very small, nearly compensated, residue moment integrated over thickness. These behaviors rule out the possibility of a strong interlayer-coupled FM state, because in that case it will favor a domain switching process where magnetic moments are fully polarized along the c-axis in individual domains but pointing along opposite directions between lateral neighbors, which would generate a strong MFM signal near the domain wall, at least stronger than that of the fully polarized state at high fields,[24] as illustrated in Fig. 1a. There could be two possible scenarios: a) The inner layers have a layered AFM state so that the remnant state has a zero moment. The interlayer AFM coupling in the inner layers should be much weaker than that in the surface layers, therefore the inner layers switch at much lower fields. b) The inner layers have an FM interlayer coupling, but domains form along the c-axis during the switching in order to lower the magnetostatic energy. In this case, the FM-coupled inner layers can still have a nearly compensated moment at remnant. The interlayer FM coupling strength should be very weak to allow a small c-axis domain size, and it also explains that the switching onsets before the applied field reverses direction because of the demagnetization field. In both cases, the magnitudes of the interlayer exchange coupling should be small, while our measurements cannot conclusively determine its sign. (See further discussion on the sign below.)

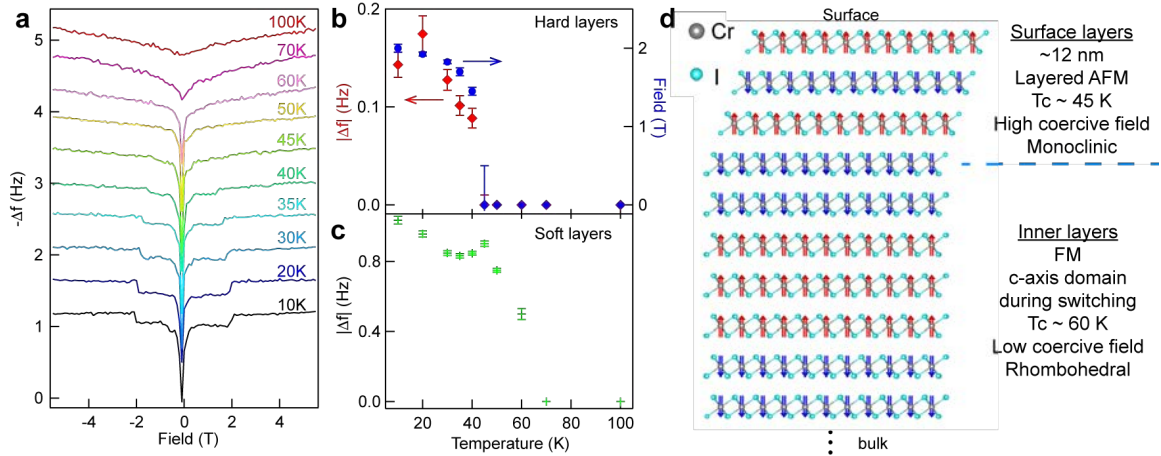

**Figure 3. Temperature dependence of MFM signals in the surface and inner layers.** (a) MFM vs magnetic field in the 200 nm CrI$_3$ flake at temperatures from 10 K to 100 K. (b) The MFM signal change and the switching field of the hard layer as a function of temperature. (c) The MFM signal associated with the switching of soft layers as a function of temperature. (d) Illustration of the stacking orders and spin configurations in surface and inner layers.

The remaining question now is what causes the magnetic properties in the surface layers to differ from the inner layers. Recent experimental[13,17–19] and theoretical[20–23] studies have uncovered a close relation between the magnetic order and the layer stacking order in CrI$_3$. Bulk

CrI$_3$ is in a monoclinic phase at room temperature and undergoes a structural phase transition to a rhombohedral phase at around 200 K.[14,25] The main difference between the two phases is different stacking orders of the CrI$_3$ layers. Second harmonic generation[13], high pressure[17,18], and Raman[19] experiments demonstrate that exfoliated few-layer CrI$_3$ flakes remain in the monoclinic phase at all temperatures. Given the similarity in magnetic properties between the surface layer group in our samples and few-layer CrI$_3$ flakes, and between the inner layer group and CrI$_3$ bulk crystals, we believe that the surface layers, which is about ~13 nm thick at each surface, should be in the monoclinic phase at all temperatures like few-layer CrI$_3$ flakes, while the inner layers undergo the structural phase transition like bulk CrI$_3$. Therefore, the monoclinic surface layers exhibit a layered AFM order with a Curie temperature of 45 K, while the rhombohedral inner layers have a higher Curie temperature at around 60-70 K with weaker interlayer coupling strengths. Exfoliated CrI$_3$ flakes thicker than ~25 nm will have these two phases coexisting, while thinner flakes are entirely in the monoclinic phase.

Recent high-pressure measurements[17,18] have demonstrated that few-layer CrI$_3$ flakes can be switched from the monoclinic to the rhombohedral phase and the magnetic order switches from the layered AFM state to the FM state accordingly. These results suggest that the rhombohedral phase of CrI$_3$ favors an FM interlayer coupling. Correlating this conclusion with our results supports the proposed scenario b), i.e., the inner layers have a weak FM coupling and their switching at low fields proceeds with formation of c-axis domains.

The mechanism by which the structural phase transition is suppressed in the surface layers of CrI$_3$ flakes is likely due to surface effect, e.g., defects in the surface layer, the interaction between the surface layer and hBN, surface strains created during the mechanical exfoliation process, or simply just having a surface where the interlayer coupling only comes from one side. Our results indicate that such surface effect can modify the stacking order not only in the topmost few layers but roughly ~13 nm (almost 20 layers) into the bulk. The stacking order in these ~20 layers should be quite uniform as indicated by the sharp switching around 2 T consistent in different samples. The switching of the inner layers exhibits more variations across samples (see, e.g., the flakes in Fig. 2a and Fig. 2b and more data in SI), which could possibly be due to a transition phase between the two different stacking orders. It would be interesting to investigate how the layer stacking transits from the monoclinic to the rhombohedral phase across their interface. The correlation between magnetic order and layer stacking has also been found in two other chromium trihalides, CrBr$_3$[26] and CrCl$_3$[27], so multiple structure phases and magnetic orders could also coexist in their exfoliated flakes. The coexistence of the two magnetic configurations naturally forms a magnetic heterostructure within the same material but with drastically different magnetic properties, the AFM phase at two surfaces and the FM phase in the interior layers. Further studies on the electronic properties across such sandwich structure could potentially realize new device functionalities, and developing methods to control the transition and thickness of the two phases would provide new opportunities to engineer magnetic states in these van der Waals materials and new insights on the magnetism in two dimensions.

**Methods**
CrI$_3$ flakes were mechanically exfoliated in an Ar-filled glovebox and a dry transfer method is used to form a sandwich structure where CrI$_3$ is protected by thin hBN flakes on both sides. (See further fabrication details in SI.) Magnetic force microscopy measurement was performed on a home-built cryogenic scanning probe platform. Commercial MFM probes with Co-Cr coating and a nominal coercivity of 400 Oe were used. In MFM measurement, the probe was scanned over the sample surface at a constant height of ~100-150 nm. The MFM signal, i.e., the change in the cantilever resonant frequency was measured using a phase locked loop. During scanning, a potential feedback loop similar to that in an electrostatic force microscope (EFM) is used to apply a tip bias to compensate the tip-sample potential difference, in order to minimize the electrostatic force on the probe. More details on the potential compensation are provided in SI.

**Acknowledgments**
We thank Kin Fai Mak, Jie Shan, and Liuyan Zhao for helpful discussions. The MFM work was supported by the start-up funds from the Regents of the University of California. T.S., J.X.L., M.L. and J.S. acknowledge support from DOE BES grant DE-FG02-07ER46351. B.N. and D.W. acknowledge support from the Natural Science Foundation of China (51725203, 51721001 and U1932115). B.N. and T.S.'s visits at UCR were supported by the China Scholarship Council (CSC). K.W. and T.T. acknowledge support from the Elemental Strategy Initiative
conducted by the MEXT, Japan and the CREST (JPMJCR15F3), JST.

**Supporting Information.**
Sample fabrication, experimental setup, MFM images during the magnetic field sweep, determination of the hard layer thickness, thickness dependence, temperature dependence

**Author Contributions**
Y.T.C. and J.S. initiated and supervised the project. B.N. performed the MFM measurements with assistance from X.H. and B.A.F.. T.S. fabricated the device with help from M.L., J.X.L., Y.X.. B.A.F. performed MFM simulations. S.G. and F.K. carried out the Raman characterization under supervision of A.B.. D.W. provided support for B.N.'s visit to UCR. T.T. and K.W. provided hBN crystals. B.N., T.S., J.S. and Y.T.C analyzed the data and wrote the paper with comments from all authors.

# Supporting Information

## Coexistence of Magnetic Orders in Two-Dimensional Magnet CrI$_3$

Ben Niu,[1,2,†] Tang Su,[3,2,†] Brian A. Francisco,[2] Subhajit Ghosh,[4] Fariborz Kargar,[4] Xiong Huang,[5] Mark Lohmann,[2] Junxue Li,[2] Yadong Xu,[2] Takashi Taniguchi,[6] Kenji Watanabe,[6] Di Wu,[1] Alexander Balandin,[4] Jing Shi,[2,*] Yong-Tao Cui[2,*]

1. National Laboratory of Solid State Microstructures, Department of Materials Science and Engineering, Jiangsu Key Laboratory for Artificial Functional Materials, and Collaborative Innovation Center of Advanced Microstructures, Nanjing University, Nanjing 210093, China
2. Department of Physics and Astronomy, University of California, Riverside, CA 92521, USA
3. International Center for Quantum Materials, School of Physics, Peking University, Beijing 100871, P. R. China
4. Department of Electrical and Computer Engineering, University of California, Riverside, CA 92521, USA
5. Department of Materials Science and Engineering, University of California, Riverside, CA 92521, USA
6. National Institute for Materials Science, 1-1 Namiki, Tsukuba, 305-0044, Japan

† These authors contribute equally to this work.
* Email: jing.shi@ucr.edu, yongtao.cui@ucr.edu

Contents:

**Section I. Sample fabrication**

CrI$_3$ flakes are exfoliated inside an argon-filled glovebox. A suitable CrI$_3$ flake is dry transferred and released either directly on top of a thin hBN flake or on a Pt (5 nm) covered hBN flake. The Pt film helps to screen the electrostatic potential from the SiO$_2$ substrate. Another hBN flake is then transferred on top of the CrI$_3$ to protect it from degradation. With the hBN capping, the sample can then be taken out of the glovebox for MFM measurement. The CrI$_3$ quality is confirmed with Raman spectroscopy (see Fig. S1).

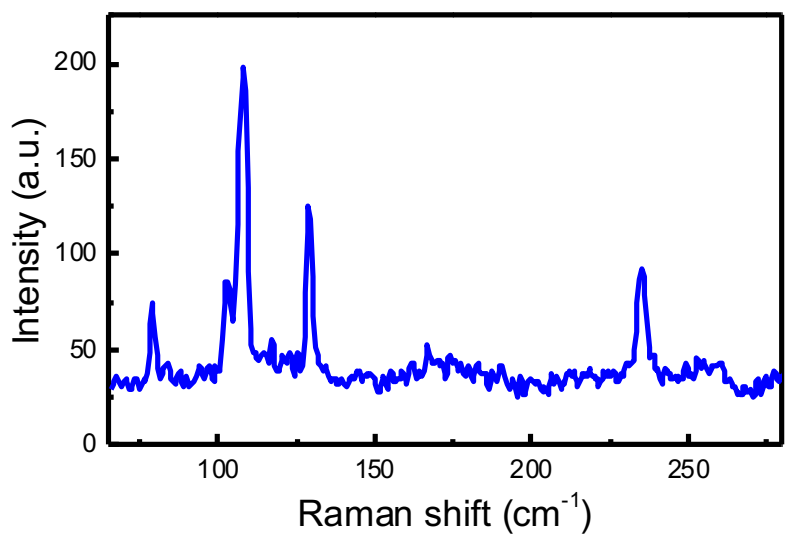

**Figure S1. A typical Raman spectrum of the CrI$_3$ flake taken at room temperature.**

**Section II. Experimental setup**

MFM measurements are performed in a home-built low temperature scanning probe microscope with Co-Cr coated commercial cantilever probes. The probe driven at its resonance (~75 kHz) by an AC excitation is raster scanned over the sample surface at constant heights to avoid topography induced artifacts. Phase locked loop (PLL) is used to obtain MFM signal, the change in the probe's mechanical resonant frequency.

1. <u>MFM measurement simultaneously with electrostatic force microscope (EFM)</u>

The change in the probe's mechanical resonance can be caused not only by the long-range magnetic force but also by the long-range electrostatic force. CrI$_3$ is a semiconductor with a bandgap of 1.2 eV, thus charge disorders possibly introduced during the fabrication process could lead to potential variation on the sample surface which will generate long-range electrostatic force. To nullify the effect of the potential disorder, we simultaneously perform EFM to compensate the electrostatic force. As illustrated in Fig. S2, at 2 T, when domain pattern appears during the reversal process, EFM assisted MFM clearly uncovers magnetic domains (Fig. S2 (b)) which were dominated by the potential disorder (dark blue regions in Fig. S2 (a)), extracting the electrical potential distribution in Fig. S2 (c).

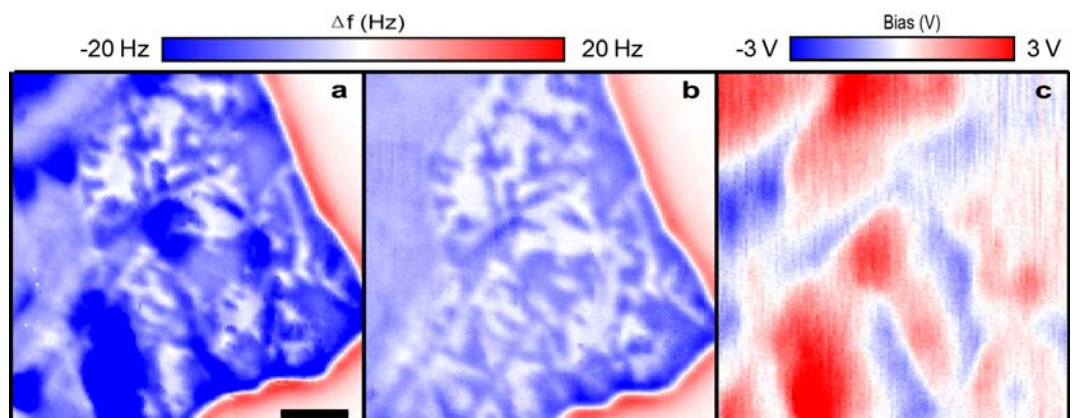

**Figure S2. Performance of EFM-assisted MFM.** MFM imaging of a 200 nm flake at 10 K and 2 T simultaneously with EFM (b) and without EFM (a). (c) demonstrates the potential distribution on the sample surface. Scale bar is 2 μm.

2. Quantitative characterization of the MFM signal as a function of magnetic field

In case of laterally uniform magnetization, MFM signal is proportional to the area density of moment integrated over the sample thickness. (See Section IV for further discussions on this property) To probe the magnetic moment more quantitatively, we perform MFM measurement in areas with fewer domains. Note that some of the contrast patterns seen in the MFM image at high field where the moments are expected to be fully polarized are due to regions of different thickness, which can be correlated well with the topography of the flake show in Fig. S3 (a). The tip is then scanned across the sample edge repeatedly along a single line as illustrated by the dashed arrow in Fig. S3 (b) while the field along c axis is varied from +5 T to -5 T. An example scan profile measured at 5 T is shown in Fig. S3 (c) in which a pair of maxima and minima indicates a dramatic change in the derivative of the magnetic field across the sample edge. There is also a peak in the sample region which is due to a topography edge between regions of different thicknesses. Away from that, the steady signal inside the sample bulk indicates a laterally uniform magnetization fully polarized by the applied field. The background signal corresponding to the non-magnetic substrate is further subtracted from each line. The resulting MFM signal is spatially averaged inside the sample bulk at each field and then plotted as a function of the field, which generates the MFM vs B plots. Importantly, sample position drifts should be corrected when sweeping the field.

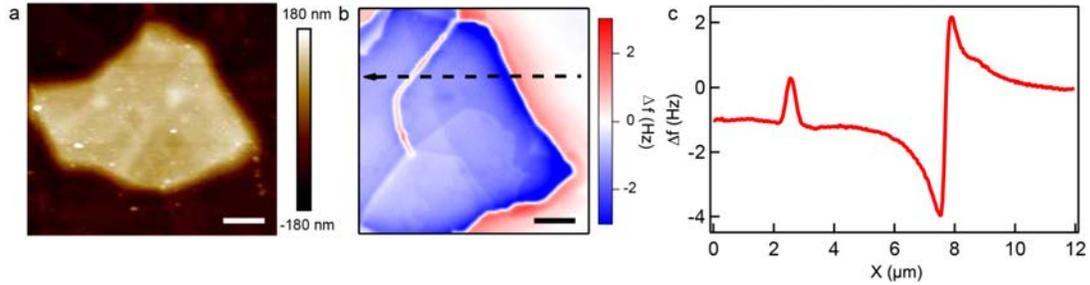

**Figure S3. Scan profile at 5 T.** (a) Topography imaging of the 200 nm flake. Scale bar is 5 μm (b) MFM imaging at 10 K and 5 T and a scanning line across the sample edge. Scale bar is 2 μm (c) Scan profile along the line. The signal difference between sample and non-magnetic substrate is extracted as the MFM signal.

**Section III. MFM images**

Figure S4 supplements MFM images of the flake in Fig. 1(b) at detailed fields above -0.1 T (Fig. S4 (a)) and below -0.1 T (Fig. S4 (b)). The domains due to the partial reversal are most pronounced at 2.05 T. Below -0.1 T, similarly to the trend of the positive field side, the average MFM signal in the bulk changes rapidly from -0.15 T to -0.7 T, then the domain configuration persists as the field is further reduced to -5 T except within a narrow range between -1.95 T and -2.1 T where magnetic domains appear.

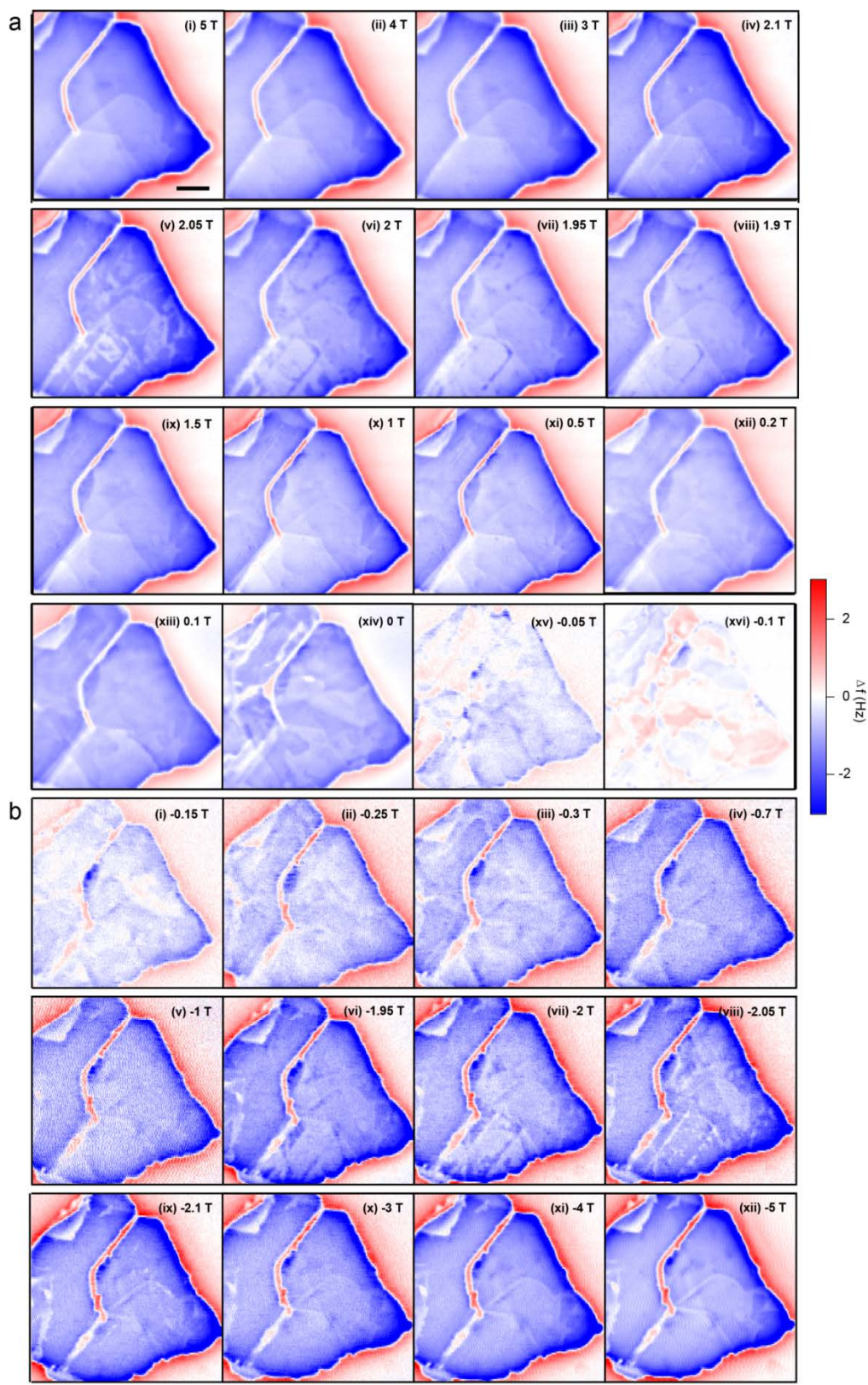

**Figure S4. Field dependent MFM images at 10 K.** At fields above (a) and below -0.1 T (b). Scale bar is 2 μm.

## Section IV. Determination of the hard layer thickness

We calculated the thickness of the hard layer by multiplying the total layer thickness with the percentage fraction of the MFM signal change at 2 T which corresponds to the hard layer switching. This method assumes that the MFM signal is proportional to the thickness of the ferromagnetic layers that exhibit a net vertical moment when it is in a uniform magnetic state, i.e., without magnetic domains. We justify this assumption with finite element simulation, as shown in Fig. S5. The simulated MFM signal is the derivative of the z-component of the magnetic field with respect to z, $dB_z/dz$, evaluated at a fixed height (150 nm) above the surface of the magnetic flake. The thickness of the fully aligned magnetic flake is varied from 0 to 1000 nm hence the net vertical moment increases linearly. In the small thickness range relevant for our analysis (<200 nm), $dB_z/dz$ is approximately linear, and it deviates from linear behavior more significantly at large thicknesses. The linear relation between MFM signal and flake thickness in the range of interest (<200 nm) indicates that surface layers and inner layers contribute approximately equally to the MFM signal, therefore the MFM signal change can be used to extract the thickness change of the magnetically aligned layers.

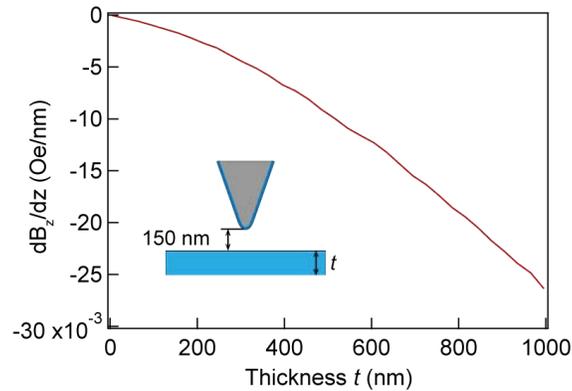

**Figure S5. Simulated $dB_z/dz$ as a function of the flake thickness.** The derivative is evaluated at a fixed height (150 nm) from the top surface of the flake.

The physics for such behavior is the following. The magnetic field configuration is sensitive to the lateral geometry of the sample. When the magnetic state is uniform in a relative wide but thin magnetic flake, the magnetic field near the flake surface decays very slowly, which applies to our case where the tip-sample distance (100 - 200 nm) is much smaller than the flake's lateral size (several μm). (Recall that in the extreme case of an infinitely large magnetic thin film, the magnetic field would be perfectly perpendicular to the film and uniform in the entire space, not decaying at all.) While $B_z$ itself already decays slowly over distance due to the large sample size relative to tip-sample distance, $dB_z/dz$ changes even more slowly near the surface. Therefore, the contributions from surface layers and inner layers at the tip location are roughly the same, leading to the linear dependence of MFM signal (proportional to $dB_z/dz$) as a function of the flake thickness. We expect such behavior will break down as the tip is moved further away from the surface. In particular, when the tip-sample distance is comparable to the lateral size of the sample (~ μm), the magnetic field will behave more like that from a dipole magnet than that from a thin film magnet.

Based on this property, the MFM signal can be used to evaluate the relative change of the magnetic moment in a thin film magnet, given the following conditions: 1) "thin film geometry": the sample thickness should be much smaller than the sample's lateral size so that the magnetic field near the sample surface decays gently; 2) "near surface probe": the tip-sample distance should also be much smaller than the sample's lateral size; 3) "uniformly magnetized state": the entire thin film should be in a uniform magnetic state without domain formation. Meanwhile, we note that the MFM signal cannot be used to obtain an absolute measurement of the magnetization because the field derivative depends sensitively on the lateral size of the sample. In our CrI3 measurement, switching of the magnetic moments in different layers only causes the effective thickness of the magnetic layer to change, and all the conditions mentioned above are satisfied. Therefore, our analysis based on the relative change of the MFM signal within the same flake provides a good estimation on the thicknesses of the hard and soft layers.

The thickness of the top hBN is in the range of 20 – 40 nm, which varies from sample to sample. The tip distance from the actual surface of the CrI3 flake will be slightly larger than the tip distance from the top hBN surface, but the total distance is still small compared to the sample size and will not change the roughly linear dependence of the MFM signal on the effective thickness of the magnetic flake.

**Section V. Thickness dependence**

Figure S6 plots MFM signal as a function of magnetic field for flakes of 110 nm, 100 nm, 80 nm, 30 nm and 25 nm at 10 K. The two-stage switching behavior is commonly observed in all samples except the 25 nm flake. The group of hard layers switches at around 2 T while the soft layers switch at low fields. Blue and red curves represent the data for increasing and decreasing the field. In general, the soft switching has a larger hysteresis than the hard switching at 2 T.

We would like to note that the absolute value of the MFM signal from samples of different lateral geometry cannot be compared directly. For example, in Fig. 2 of the main text, the MFM signal in the 200 nm flake is actually slightly smaller than that from the 110 nm flake, which is due to their different lateral size. As discussed in Section IV, the derivative of the magnetic field, $dB_z/dz$, is very sensitive to the lateral geometry of a thin flake. The 200 nm flake is ~17 μm wide while the 110 nm flake is ~6 μm wide (Fig. S7a). The larger width of the 200 nm flake will decrease the MFM signal inside the flake. As an illustration, we simulated the magnetic field derivative, $dB_z/dz$, in both a 17 μm wide and 6 μm wide stripe (the other side is infinitely long). The value of $dB_z/dz$ at 150 nm height from the flake surface is plotted in Fig. S7b. The signal in the 17 μm wide / 200 nm thick flake is almost 10 times lower than the 6 μm wide / 110 nm thick flake. In reality, considering the irregular shape of the flakes and the fact that the MFM tip has a thin magnetic coating over the entire tip body (not only at the tip apex), the measured MFM signal will be a complicated convolution of the d$B_z$/d$z$ values in space. As a result, the absolute value of the MFM signal cannot be compared across samples of different lateral geometry.

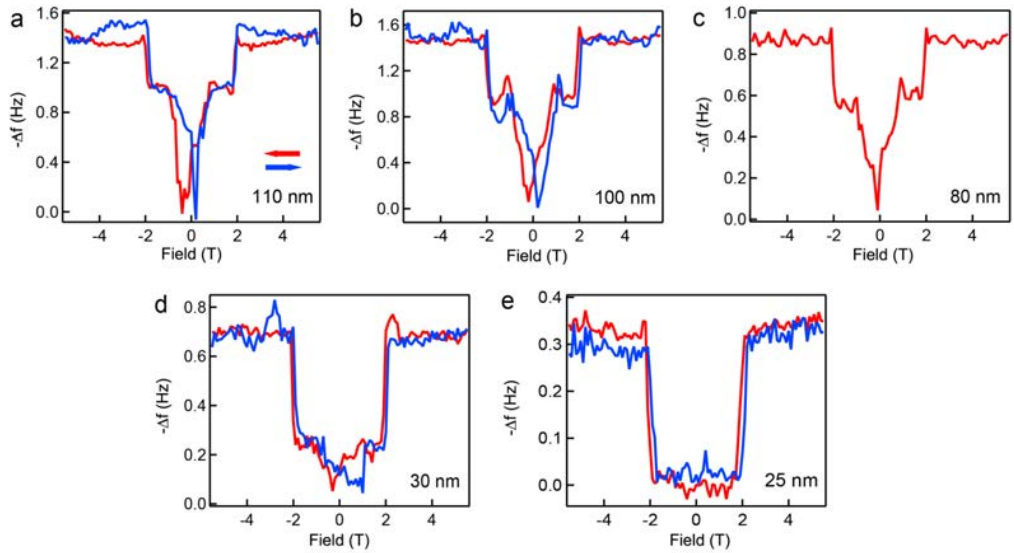

**Figure S6. Two groups coexist in samples of various thicknesses.** (a) 110 nm (b) 100 nm (c) 80 nm (d) 30 nm (e) 25 nm. Blue and red arrows indicate the direction of increasing and decreasing field.

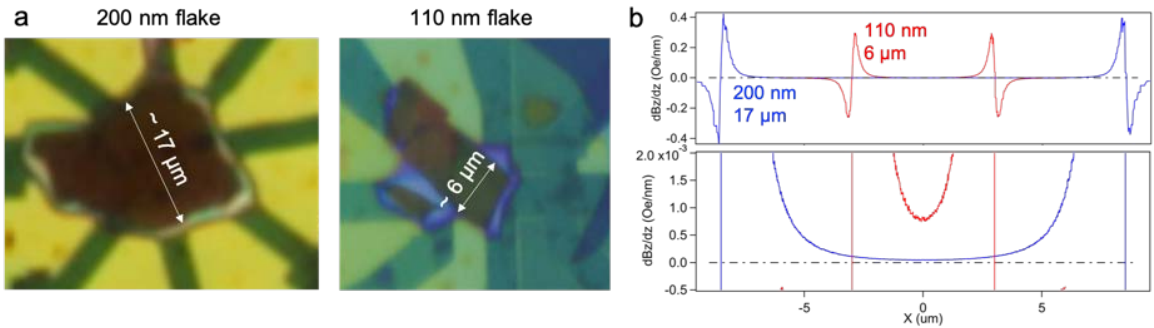

**Figure S7. Simulated MFM signal in 200 nm and 110 nm flakes** (a) Optical images of the 200 nm flake and 110 nm flake. (b) Simulated d$B_z$/d$z$ signals across a 17 μm wide / 200 nm thick (blue) and 6 μm wide / 110 nm thick (red) flake, evaluated at 150 nm from the top surface.

**Section VI. Temperature dependence**

For samples of various thicknesses, MFM signal vs B curves at temperature from 10 K to 70 K are plotted in Figure S8. Through all thicknesses, the hard switching vanishes at ~45 K while the soft switching persists up to 60 K, resembling the behaviors in CrI$_3$ thin layers and bulk crystals, respectively.

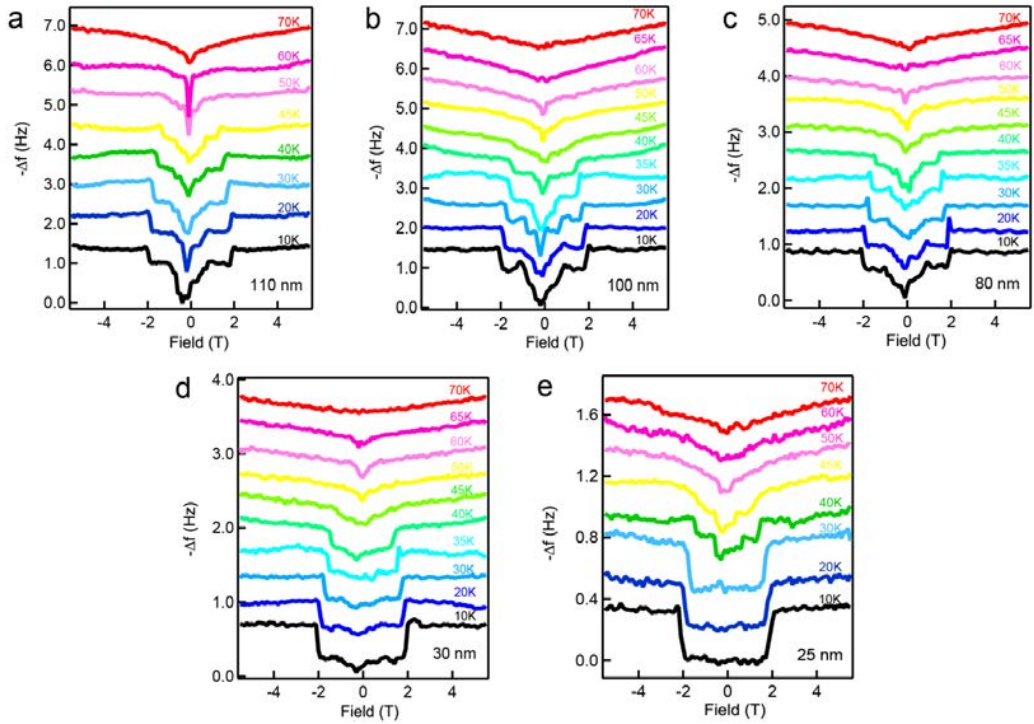

**Figure S8. Temperature dependence of MFM vs B curves in samples of various thicknesses.** (a) 110 nm, (b) 100 nm, (c) 80 nm, (d) 30 m, (e) 25 nm.